# Performance Analysis of MANET Routing Protocols Using An Elegant Visual Simulation Tool

Nazmus Saquib[1], Md. Sabbir Rahman Sakib[1], and Al-Sakib Khan Pathan[2]
[1]Department of Electrical and Electronic Engineering, BRAC University
[2]Department of Computer Science and Engineering, BRAC University
66 Mohakhali, Dhaka 1212, Bangladesh
{nsaquib, srsakib}@bracu.ac.bd, sakib.pathan@gmail.com

## ABSTRACT

The task of simulation is often complicated for which many naive users often seek for relatively easier solutions. In many cases, simulations are done without any visual output which makes them non-attractive. In this paper, we present ViSim; a new simulation tool that has a user-friendly graphical interface. ViSim could be useful for researchers, students, teachers in their works, and for the demonstration of various wireless network scenarios on computer screen. It could make the task of simulation more exciting and enhance the interest of the users without going into complex command-only text interface. ViSim is not a simulation engine rather it calls ns-2 simulations in the background and makes the task easy for the users to visualize the simulation in Windows environment. Though ViSim is mainly a simulation demonstration tool, any user with the knowledge of ns-2 and Tcl scripting is also allowed to do necessary modifications and quick configurations for any other MANET routing scenario. Using our simulation tool, we measured the performances of several Mobile Ad-hoc Network (MANET) routing protocols. In this paper, we present the performance analysis of three prominent MANET routing protocols; DSDV, DSR, and AODV using our tool. The details of various features of ViSim, brief descriptions of the selected routing protocols and their comparisons, details about the performed experiments, and the gained results are presented in this work.

# 1 INTRODUCTION

In this modern era, the communications technology has advanced at a tremendous pace. The growing trend of using wireless communications capability built into common accessories such as laptops, PDAs, mobile phones provides us with a sea of nodes that currently lies at best not utilized to its fullest and at worst constantly fighting for the sliver of frequency spectrum that is being shared by multiple providers. Wireless ad-hoc routing protocols show promise in bringing the order and harmony to this chaotic situation.

A Mobile Ad-hoc Wireless Network (MANET) is a collection of autonomous nodes that communicate with each other by forming a multi-hop network, maintaining connectivity in a decentralized manner [1], [2]. It consists of a set of mobile hosts communicating amongst themselves using wireless links, without the use of any other communication support facilities such as base-stations. The nodes in a MANET can be PDAs, laptops or any other device that is capable of transmitting and receiving information. Each node in such a network acts as a host or end system (transmitting and receiving data) as well as a router at the same time. The nodes in a MANET are generally mobile and may go out of range of other nodes in the network. Therefore, Routing in MANET is difficult since mobility causes frequent network topology changes and requires more robust and flexible mechanism to search for and maintain the routes. When the network nodes move, the established paths may break and the routing protocols must dynamically search for other feasible routes. With a changing topology, even maintaining connectivity is very difficult. In addition, keeping the routes loop free is more difficult when the hosts move. Besides handling the topology changes, routing protocols in MANET must deal with other constraints, such as low bandwidth, limited energy consumption, and high error rates; all of which may be inherent in the wireless environment. Furthermore, the possibility of asymmetric links caused by different power levels among mobile hosts and other factors such as terrain conditions make routing protocols more complicated.

Because of these challenging features of MANET, it has been under tremendous scrutiny and interest from the time of its emergence and by this time, numerous works have tried to address various issues.

Routing in MANET is one of the well-addressed topics. Though many routing protocols have already been proposed and well-accepted in the research community because of their given promise and performance, there still remains the necessity of a flexible, user-friendly simulation tool that can make the task of simulation and visualization of routing protocols easier. Many simulators can successfully simulate various routing protocols of MANET. However, there are only a few tools to handle the simulations with a graphical interface. This particular fact has motivated us to design and develop a simulation tool so that the users could be able to deal with complex simulation scenarios in a much easier way without getting involved into the command-only interface.

We have named our simulation tool 'ViSim'. Currently the version 1.0 (ViSim 1.0) is available. We will talk in detail about our tool later in this paper. ViSim could help a network administrator to choose a particular ad-hoc routing protocol for a specific scenario through analyzing the graphs for different routing protocols. Alongside describing various aspects and features of ViSim, we have also analyzed the performances of the prominent routing protocols for MANET using our tool. We have also discussed the features of the routing protocols so that the readers could get some idea how they operate in a MANET setting.

The rest of the paper is organized as follows: Section 2 gives a brief description of three major MANET routing protocols – DSDV, DSR and AODV that we have used for our performance analysis. Section 3 describes ViSim that we have developed to investigate various simulation scenarios. After doing the performance evaluations, the obtained results about Throughput, Goodput, and Routing Loads are presented in Section 4. Section 5 talks about some relevant works and comparative features of ViSim and other tools, and finally Section 6 concludes the paper mentioning the achievements from this work with future directions for research and objectives.

## 2 ANALYZED MANET ROUTING PROTOCOLS

We have analyzed three well known routing protocols for MANET. Before discussing our performance analysis and results, in this section we briefly mention the operational methods and major features of these protocols.

**2.1 Dynamic Destination-Sequenced Distance-Vector Routing Protocol (DSDV)**

DSDV [3] is a proactive routing protocol which is developed on the basis of Bellman-Ford routing [4] algorithm with some modifications. In this protocol, each mobile node in the network keeps a routing table. Each of the routing table contains the list of all available destinations and the number of hops to each. Each table entry is tagged with a sequence number which is originated by the destination node. Periodic transmissions of updates of the routing tables help maintaining the topology information of the network. If there is any new significant change for the routing information, the updates are transmitted immediately. So, the routing information updates might either be periodic or event-driven. DSDV protocol requires each mobile node in the network to advertise its own routing table to its current neighbors. The advertisement is done either by broadcasting or by multicasting. By the advertisements, the neighboring nodes can know about any change that has occurred in the network due to the movements of nodes.

The routing updates could be sent in two ways: one is called a 'full dump' and another is 'incremental'. In case of full dump, the entire routing table is sent to the neighbors whereas in case of incremental update, only the entries that require changes are sent. Full dump is transmitted relatively infrequently when no movement of nodes occur. The incremental updates could be more appropriate when the network is relatively stable so that extra traffic could be avoided. But, when the movements of nodes become frequent, the sizes of the incremental updates become large and approach the network protocol data unit (NPDU). Hence, in such a case, full dump could be used. Each of the route update packets also has a sequence number assigned by the transmitter. For updating the routing information in a node, the update packet with the highest sequence number is used as the highest number means the most recent update packet. Each node waits up to certain time interval to transmit the advertisement message to its neighbors so that the latest information with better route to a destination could be informed to the neighbors.

## 2.2 Dynamic Source Routing (DSR)

DSR [5] is a reactive routing protocol which allows nodes in the MANET to dynamically discover a source route across multiple network hops to any destination. In this protocol, the mobile nodes are required to maintain *route caches* or the known routes. The route cache is updated when any new route is known for a particular entry in the route cache.

Routing in DSR is done using two phases: route discovery and route maintenance. When a source node wants to send a packet to a destination, it first consults its route cache to determine whether it already knows about any route to the destination or not. If already there is an entry for that destination, the source uses that to send the packet. If not, it initiates a route request broadcast. This request includes the destination address, source address, and a unique identification number. Each intermediate node checks whether it knows about the destination or not. If the intermediate node doesn't know about the destination, it again forwards the packet and eventually this reaches the destination. A node processes the route request packet only if it has not previously processed the packet and its address is not present in the route record of the packet. A route reply is generated by the destination or by any of the intermediate nodes when it knows about how to reach the destination.

## 2.3 Ad Hoc On-Demand Distance Vector Routing (AODV)

AODV [6] is basically an improvement of Dynamic Destination-Sequenced Distance-Vector (DSDV) routing protocol. But, AODV is a reactive routing protocol instead of being proactive. It minimizes the number of broadcasts by creating routes based on demand, which is not the case for DSDV. When any source node wants to send a packet to a destination, it broadcasts a route request (RREQ) packet. The neighboring nodes in turn broadcast the packet to their neighbors and the process continues until the packet reaches the destination. During the process of forwarding the route request, intermediate nodes record the address of the neighbor from which the first copy of the broadcast packet is received. This record is stored in their route tables, which helps for establishing a reverse path. If additional copies of the same RREQ are later received, these packets are discarded. The reply is sent using the reverse path. For route maintenance, when a source node moves, it can reinitiate a route discovery process. If any intermediate node moves within a particular route, the neighbor of the drifted node can detect the link failure and sends a link failure notification to its upstream neighbor. This process continues until the failure notification reaches the source node. Based on the received information, the source might decide to reinitiate the route discovery phase.

Each of these routing protocols was tested in our simulation experiments. The details of our works and our built prototype are mentioned in the following sections.

## 3 VISIM: A VISUAL SIMULATION TOOL

### 3.1 Building Blocks of ViSim

We have used two software in Windows environment for our work; ActiveTcl and Microsoft Visual Basic. Before describing ViSim's features and functionalities, in this sub-section we talk about these briefly.

ActiveTcl is an industry-standard Tcl distribution, available for Windows, Linux, Mac OS X, Solaris, AIX and HP-UX. This software creates an environment in Windows to run the ns-2 [7] simulations and .tcl scripts. It is capable of executing the simulation faster than cygwin [8]. This package contains ns.exe and nam.exe, two executable files. Once a .tcl script is written referring to a particular scenario with specifications of different simulation parameters such as ad hoc routing protocol name, number of nodes, nodal positions, MAC layer type, simulation area, time, etc; ActiveTcl can run the simulation in Windows environment. For details of ActiveTcl and its latest versions, the readers are encouraged to visit the URL; http://www.activestate.com/activetcl/ .

Microsoft Visual Basic is a popular software that we have used for developing ViSim prototype so that it can connect the simulation related tasks with a user-friendly graphical interface. For our work, we have used ActiveTcl8.3.5 and Visual Basic 6.0.

## 3.2 Overview of ViSim

Our graphical simulation tool, ViSim is built using Visual Basic 6.0 in order to make comparisons among various MANET routing protocols since there are very few prototypes available today for performing such type of task. Most of the available tools are somewhat not user-friendly. Hence, keeping that in mind, we built ViSim in such a way that any naive user can also be able to use this tool to visualize the background simulations done in ns-2 (that is run with the help of ActiveTcl in Windows operating system). ViSim runs associated .tcl files for all the three mentioned protocols (DSDV, DSR, AODV) and extracts the required information from the trace files that are generated. Eventually the graphs are plotted for different performance indicators such as Throughput, Goodput, and Routing Loads. ViSim can make the task of a network administrator easy to decide which routing protocol would be better for a particular MANET scenario.

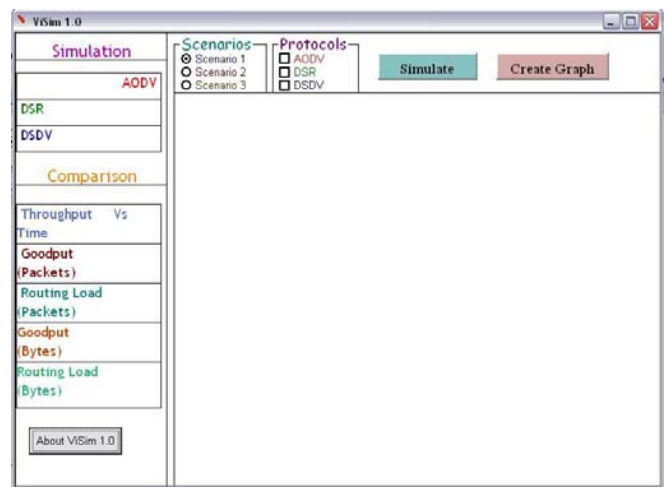

**Figure 1. Visual Simulation Tool Interface, ViSim 1.0 (run in Windows XP)**

## 3.3 Different Working Areas

Figure 1 shows the ViSim prototype/tool when it is run in Windows environment for the first time. The graphical interface has some working areas and functionalities that should be known before using it for analysis of various parameters. There are mainly four portions/areas on the ViSim interface:

(a) **Simulation:** In this area, three routing protocols are mentioned. Clicking on the names of each protocol gives the options of simulating three network scenarios using that particular protocol. The network scenarios could be modified as required in the .tcl scripts that run in the background.

(b) **Comparison:** This area has the options; Throughput vs Time, Goodput (Packets), Routing Load (Packets), Goodput (Bytes), and Routing Load (Bytes). All these buttons are used to select the parameters that the user needs for the performance analysis and comparison among the routing protocols.

(c) **Scenarios and Protocols:** This area specifies the options of three network scenarios (radio buttons) and three routing protocols (tick boxes). Also it has two buttons namely; 'Simulate' and 'Create Graph'. 'Simulate' button is used for playing the simulations and 'Create Graph' is used to plot the comparison graphs.

(d) **Output:** Output area is the right-bottom area which is shown as a blank window area when ViSim is run for the first time. Based on the choice of various options, the outputs or further options are shown in this area. The graphs are also plotted on this area when the user chooses the option of creating graphs after performing various simulations and comparisons.

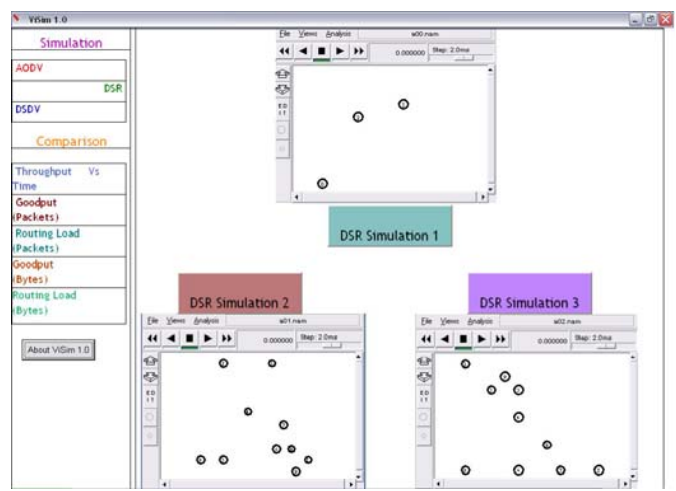

**Figure 2. DSR simulation options**

## 3.4 Functionalities of ViSim with Examples

Now, let us see the functionalities of ViSim with some practical examples. Let us suppose that we want to visualize the simulation for DSR for a particular network scenario. For this task, first we have to click DSR button under simulation area. After clicking DSR button, ViSim shows three more options (DSR Simulation 1, DSR Simulation 2, and DSR Simulation 3) on the output area as shown in Figure 2.

From these three options any one could be chosen. For our task, let us choose DSR Simulation 3. After clicking this button, ViSim calls ns-2 in its background, then reads .tcl file that specifies the simulation scenario 3, generates .nam and .tr files. Once the .nam and .tr files are generated, ViSim calls the NAM (Network Animator tool) in its background and reads the generated .nam files. Consequently, it shows a display for visual simulation [see Figure 3].

On the NAM screen, there are few buttons such as play, forward, backward, stop buttons available to control the simulation as these are done usually in Linux based environment with ns-2 and NAM. To see the visual simulation on the screen, the play button should be clicked. Like any other simulation using NAM, we can also change the step size of the simulation.

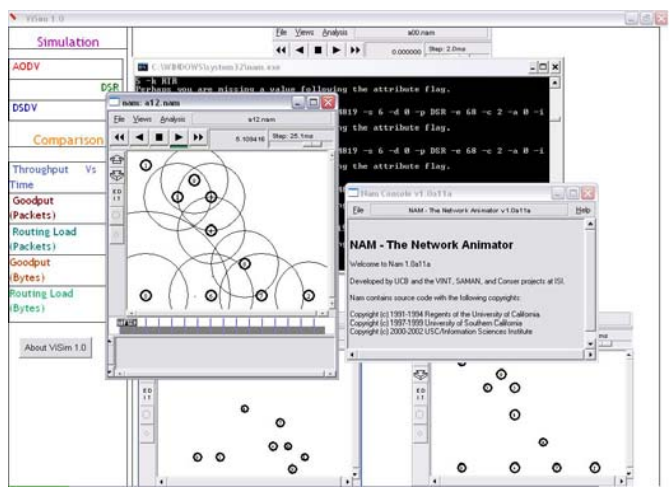

**Figure 3. The output after choosing DSR Simulation 3**

Now, if we want to make comparisons among three different protocols for performance analysis, we have to choose a specific network scenario. In our case, let us select Scenario 1. Then we have to select three mentioned protocols (or, any two or one) and side by side the performance indicators should be clicked from the five options in the comparison area. Figure 4 shows the output where we selected 'Throughput Vs Time'.

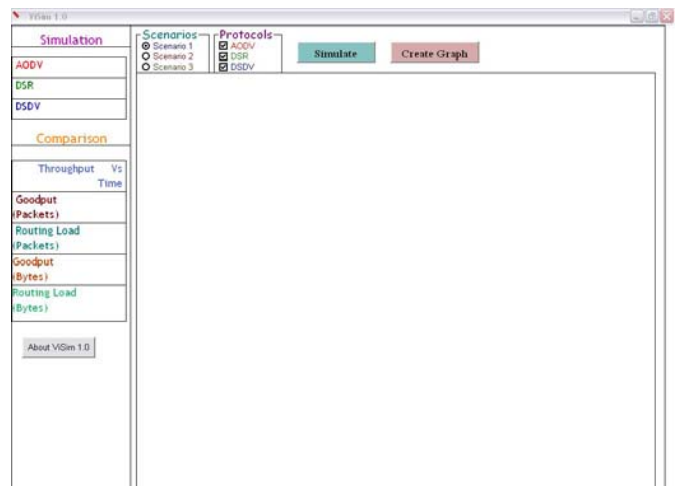

**Figure 4. An example where 'Throughput vs Time', Scenario 1, and three protocols are selected for running the simulations**

Once the simulations are performed by clicking the 'Simulate' button, we can use the generated results in the background for plotting comparison graphs. Basically, this 'Simulate' button facilitates performing various simulations with three protocols for a particular network scenario at the same time. This reduces the burden of doing the tasks repeatedly or selecting one protocol at a time under Simulation area. Once all the simulations are completed, the graph can be generated by clicking the 'Create Graph' button. By clicking 'Create Graph' button, we send the command to read the generated .tr files (trace files) and extract the required information/values from those. These values are used to plot the graphs for different protocols for a specific scenario and for different performance indicators. Figure 5 shows a sample output of what we have done so far (as an important note it should be mentioned that each simulation and plotting of graph takes a bit time as required by ns-2; for example in our case, it took about 25 seconds to plot the graph on the ViSim output area).

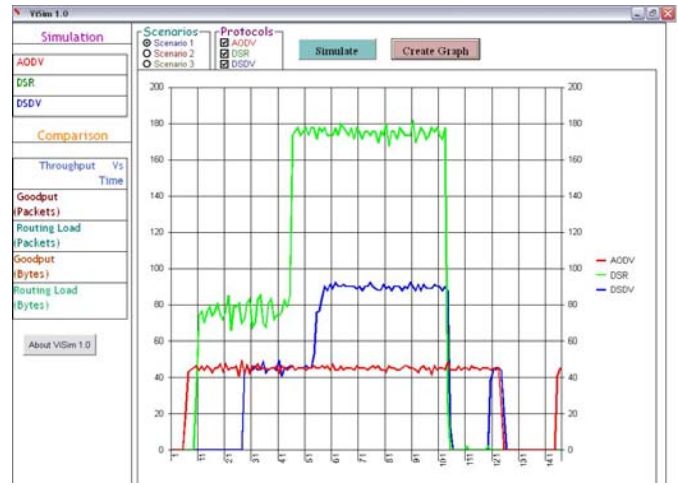

**Figure 5. A sample output graph (Throughput vs Time) using network scenario 1, all three protocols; DSDV, DSR, and AODV are compared**

Let us talk about the working mechanism of ViSim buttons a bit. When the user selects the simulation option in order to view the simulation for a particular scenario corresponding to the selected ad hoc network protocol, ViSim calls up a .bat file which contains shell script. This shell script calls the ns-2 and feeds files or file having extension .tcl, according to the choice of simulation. Then ns-2 generates trace file (extension .tr) and nam file (extension .nam). After that NAM is called via shell script and using NAM the script feeds .nam file into NAM which gives a GUI (Graphical User Interface) popup and using it, a user can actually observe the simulation. Again, when the user selects the Comparison option and clicks Create Graph after performing simulations, ViSim gathers the .tr files according to the choice of protocol, reads those and according to the performance indicators, it filters the data and picks up important information to generate the graph.

For ViSim, we have used some given network specifications. Note that any specification can be modified in the .tcl files according to the requirements to simulate another network setting. Also, various parameters used in ViSim code could be given new values. We are planning to make ViSim an open-source tool so that it could be customized to fit a particular wireless network (MANET, Wireless Sensor Network, etc.) scenario.

## 3.5 ViSim Operational Flow

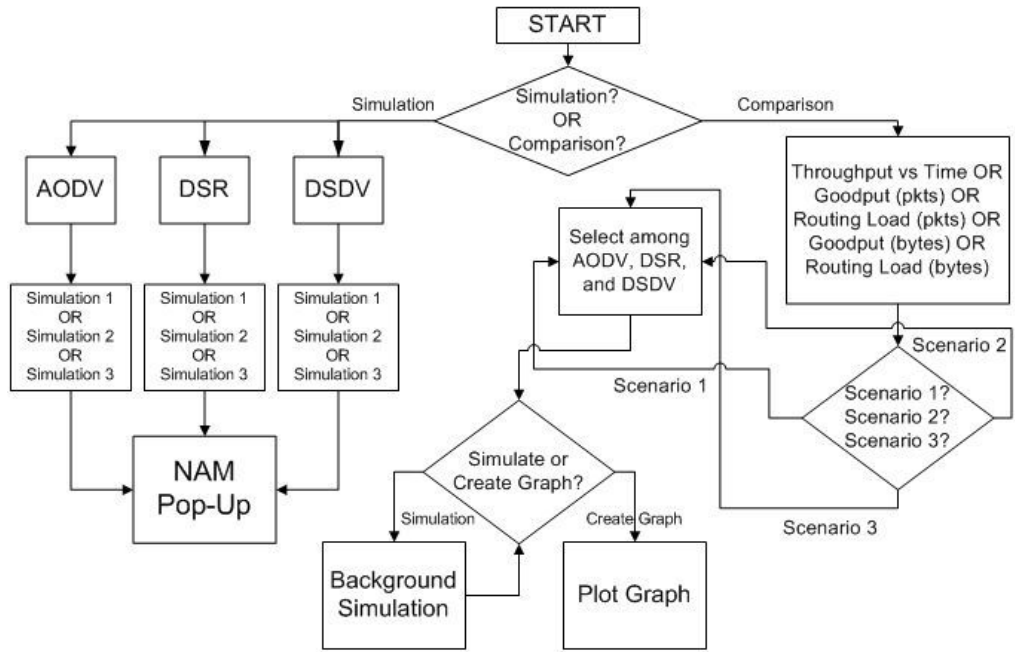

**Figure 6. Diagram of the operational flows of ViSim**

A diagram of the operational flows of ViSim is presented in Figure 6. This figure shows that after running ViSim, a user first needs to choose between two options: Simulation or Comparison. For simulation, the user is given the options of three simulations (with three different network scenarios) for each of the three routing protocols: AODV, DSR, and DSDV. Choosing any option of any simulation scenario calls NAM and then the visual simulation can be seen. On the other hand, choosing comparison gives five comparison options, then the user needs to select a particular scenario and click any one, two, or three routing protocols to compare. After that the user is given the option of simulation of all the selected items at a time or plotting of comparison graph. If there is no simulation done before plotting graph using

'Create Graph' option, it returns nothing and the output area shows blank space as usual; otherwise, if the simulations are performed and then the 'Create Graph' option is chosen, the relevant graph is plotted on the output area.

**3.6 ViSim File Organization and Simulation Scenario Modification**

In this sub-section we briefly talk about the file organization that is used in our tool. After installing ViSim a particular folder appears in which several .tcl files are kept. For any .tcl file, the general format that is used is: [The naming convention: a + Ad-hoc protocol (X) + Scenario (Y).tcl]=axy.tcl, where x represents the protocol whose value ranges from 0 to 2 and y represents the scenario whose value also ranges from 0 to 2. Now, every value of x or y is associated with the file name. Let us have a look at Table 1 for a clear idea regarding the naming convention that is used.

**Table 1. File Naming Convention ([row+column] values in the cells)**

|  | 0 (Scenario 1) | 1 (Scenario 2) | 2 (Scenario 3) |
|---|---|---|---|
| 0 (AODV) | 00 | 01 | 02 |
| 1 (DSR) | 10 | 11 | 12 |
| 2 (DSDV) | 20 | 21 | 22 |

From the above table, we may draw the following conclusions regarding the selection of a particular .tcl file for working:

a00 = "AODV Simulation for scenario 1"

a01 = "AODV Simulation for scenario 2"

a02 = "AODV Simulation for scenario 3"

a10 = "DSR Simulation for scenario 1"

a11 = "DSR Simulation for scenario 2"

a12 = "DSR Simulation for scenario 3"

a20 = "DSDV Simulation for scenario 1"

a21 = "DSDV Simulation for scenario 2"

a22 = "DSDV Simulation for scenario 3"

Therefore, if a user needs to change a file's specifications, say for example; the 2nd scenario of DSR routing protocol, he needs to open the a11.tcl file and do necessary changes. The user's manual is also available in ViSim website (mentioned later) which tells more about these facilities. We have noted this information here to indicate that ViSim could be used both as a simulation demonstration tool and a simulator tool for MANET routing protocols.

## 4 PERFORMANCE EVALUATIONS AND RESULTS

In this section, we present the obtained results from our performed experiments using ViSim tool.

### 4.1 Simulation Parameters and Specifications

Table 2 shows the specifications and parameters that we used for our experiments:

**Table 2. Simulation Parameters**

| Simulation Parameter | Value |
| --- | --- |
| Channel Type | Wireless Channel |
| Radio-propagation model | Two Ray Ground Model |
| Network interface type | Wireless Physical |
| MAC type | 802_11 b |
| Interface Queue Type | Drop Tail Primary Queue |
| Antenna model | Omni Direction |
| Number of Mobile nodes | 3-10 |
| Ad Hoc Routing Protocol | DSDV, DSR, AODV |
| Simulation Area | 500m x 400m |
| Simulation Time | 150 ms |
| Traffic Type | TCP |
| Nodal speed | 3-10 m/s |
| Packet size | 1040 Byte (Data Packets) |
|                                      | 40 bytes(Acknowledgement Packets)    |
|                                      | 60 Bytes (Routing Packets)           |
| Total Number of different Scenarios  | 15                                   |

To evaluate the performances of various routing protocols we took three network scenarios; Scenario 1 with 3 nodes, Scenario 2 and 3 with 10 nodes with different mobility characteristics. Comparisons among different protocols were based on the aggregate of the performance metrics resulting from the simulations of 15 different scenarios that had been performed for each protocol separately. To measure the performances, we used the following metrics:

**Throughput:** The total bytes received by the destination node per second (Data packets and Overhead).

**Goodput:**

*Goodput (In terms of Number of Packets):*

The ratio of the total number of data packets that are sent from the source to the total number of packets that are transmitted within the network to reach the destination.

*Goodput (In terms of Packet Size in Bytes):*

The ratio of the total bytes of data that are sent from the source to the total bytes that are transmitted within the network to reach the destination.

**Routing Load:**

*Routing Load (In terms of Number of Packets):*

The ratio of the total number of routing packets that are sent within the network to the total number packets that are transmitted within the network to reach the destination.

*Routing Load ( In terms of Packet Size in Bytes):*

The ratio of the total bytes of routing packets that are sent within the network to the total number of bytes that are transmitted within the network to reach the destination.

## 4.2 Simulation Results

To illustrate our experimental results using our tool, we first present all the outputs of 15 different cases in the Figure 7, Figure 8, Figure 9, Figure 10, and Figure 11.

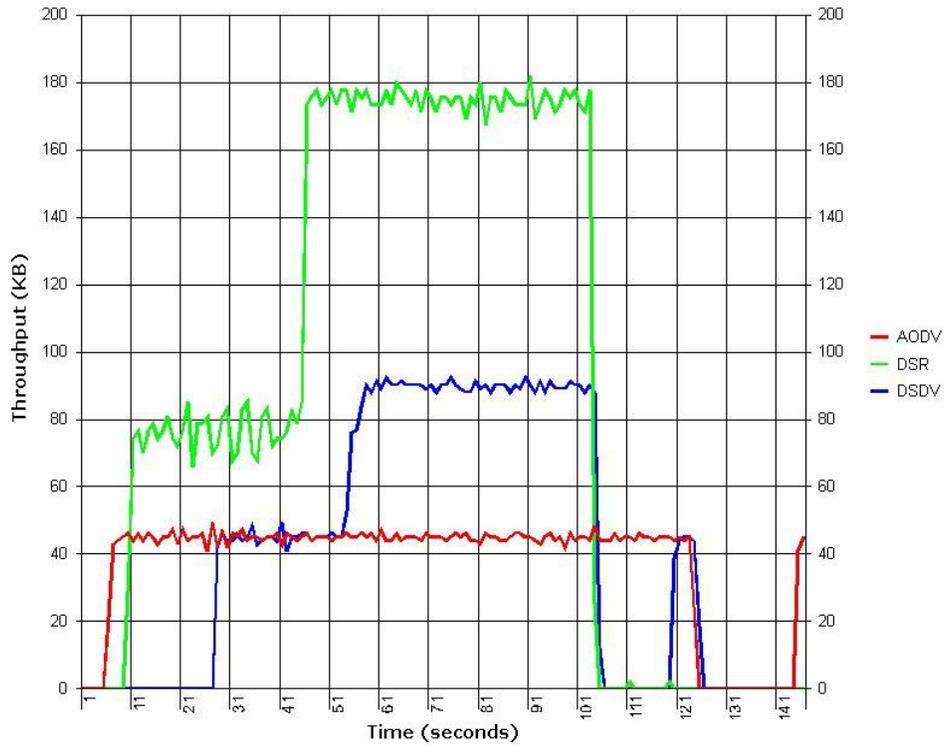

(a)

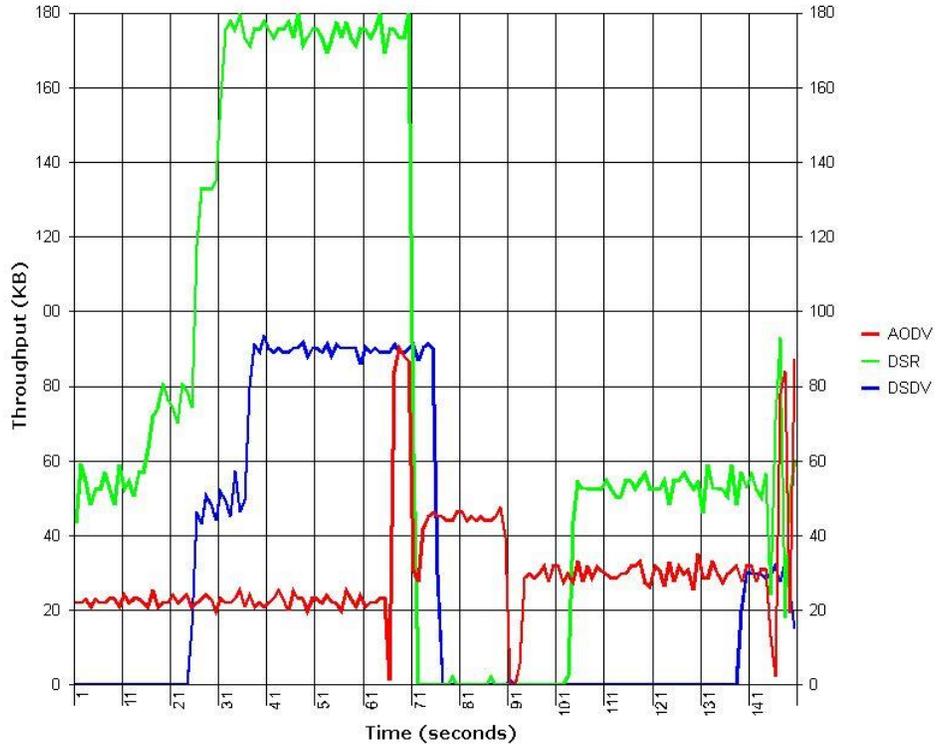

(b)

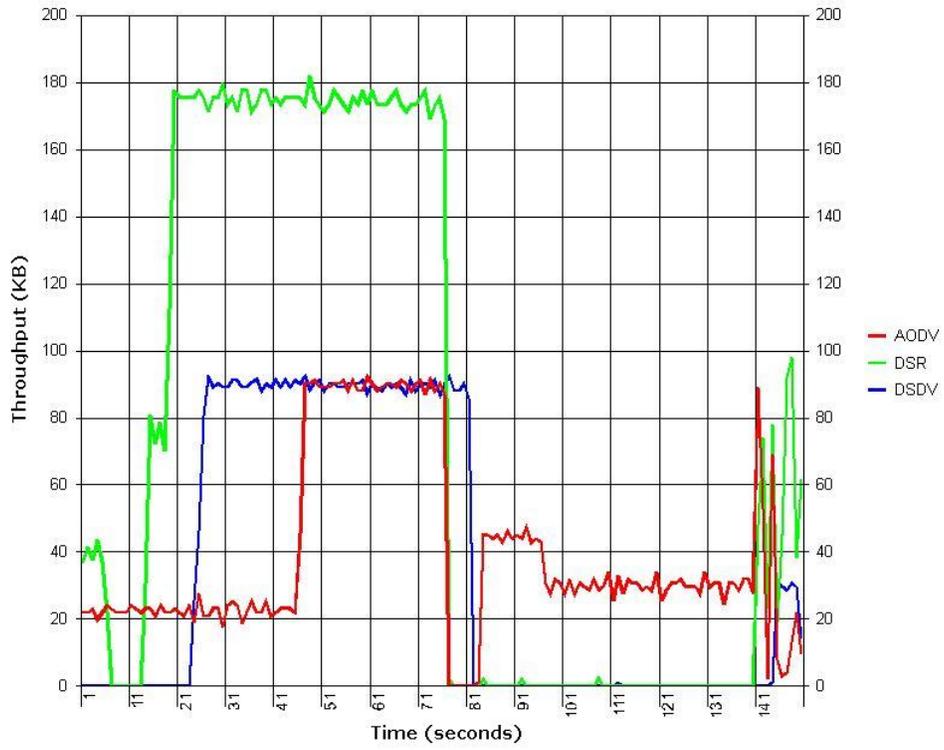

(c)

**Figure 7. Throughput vs Time (a) Scenario 1 (b) Scenario 2 (c) Scenario 3 [y axis represents Throughput (KB) and x axis represents Time (seconds)]**

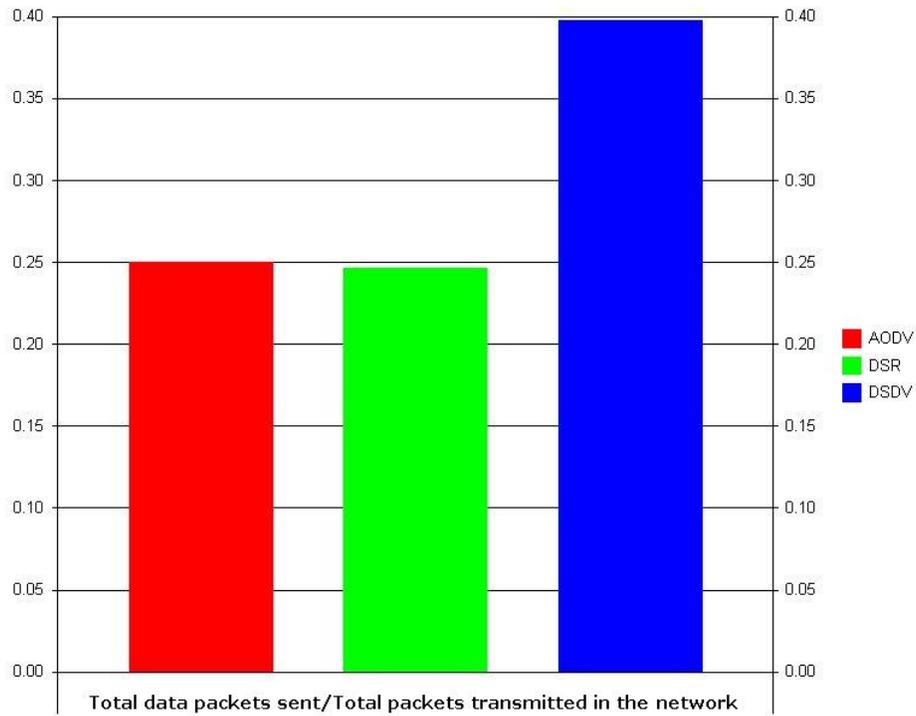

(a)

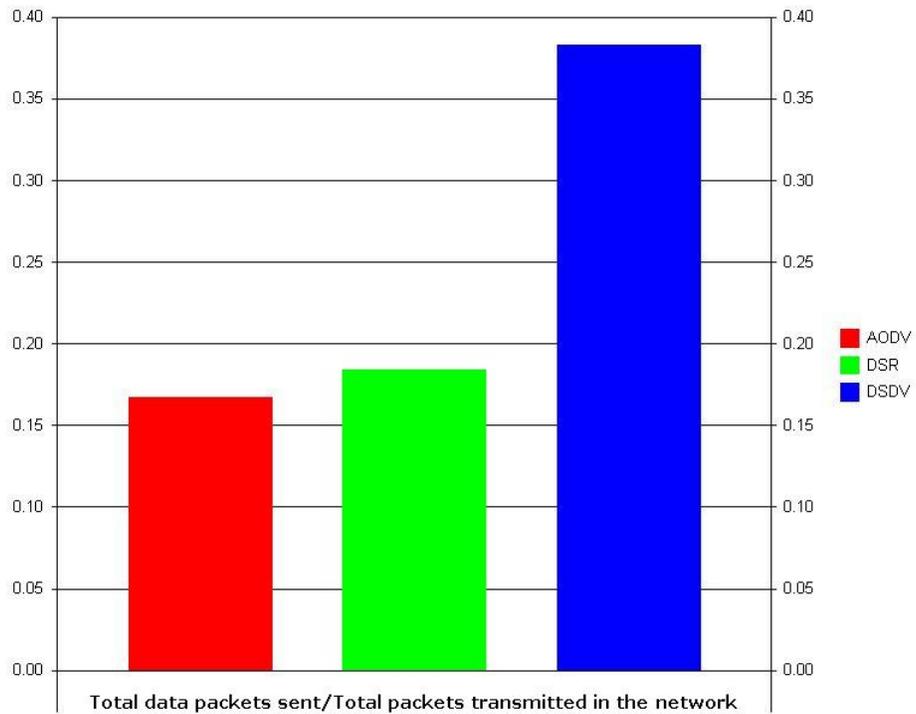

(b)

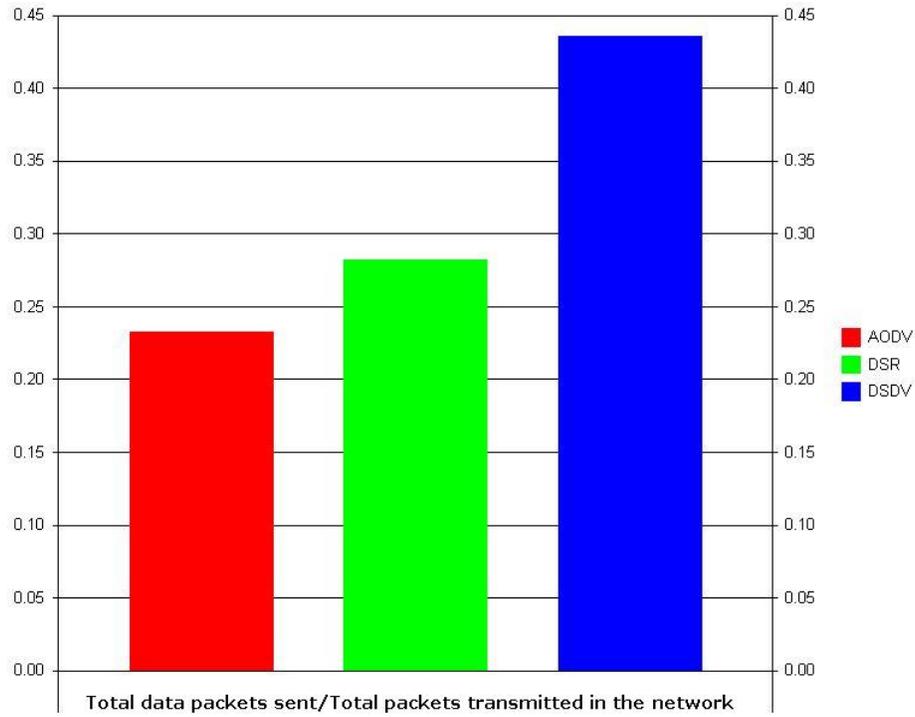

**(c)**
**Figure 8. Goodput (packets) (a) Scenario 1 (b) Scenario 2 (c) Scenario 3**

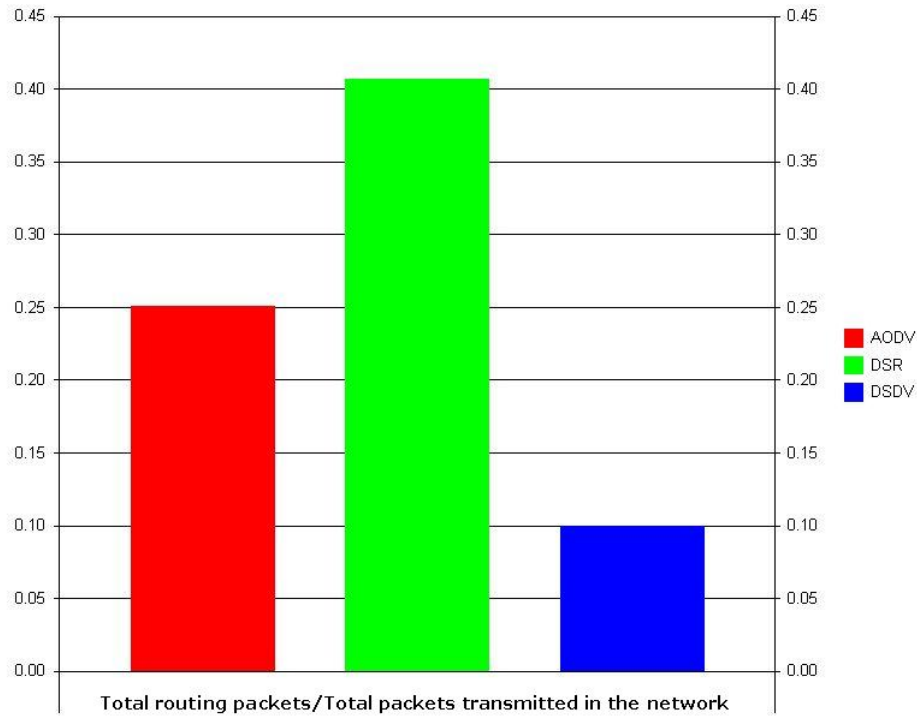

**(a)**

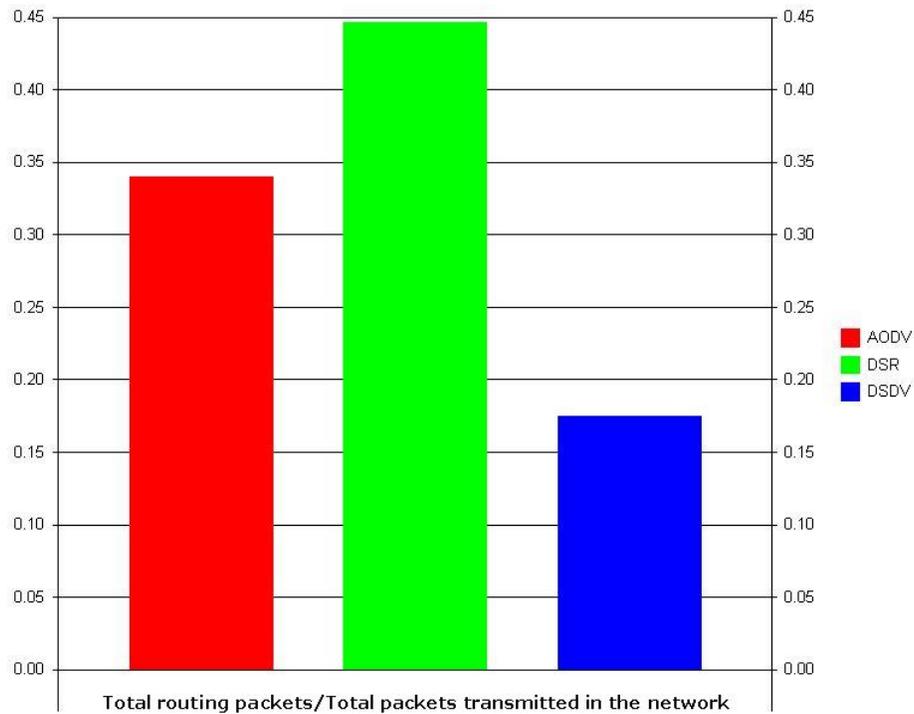

**(b)**

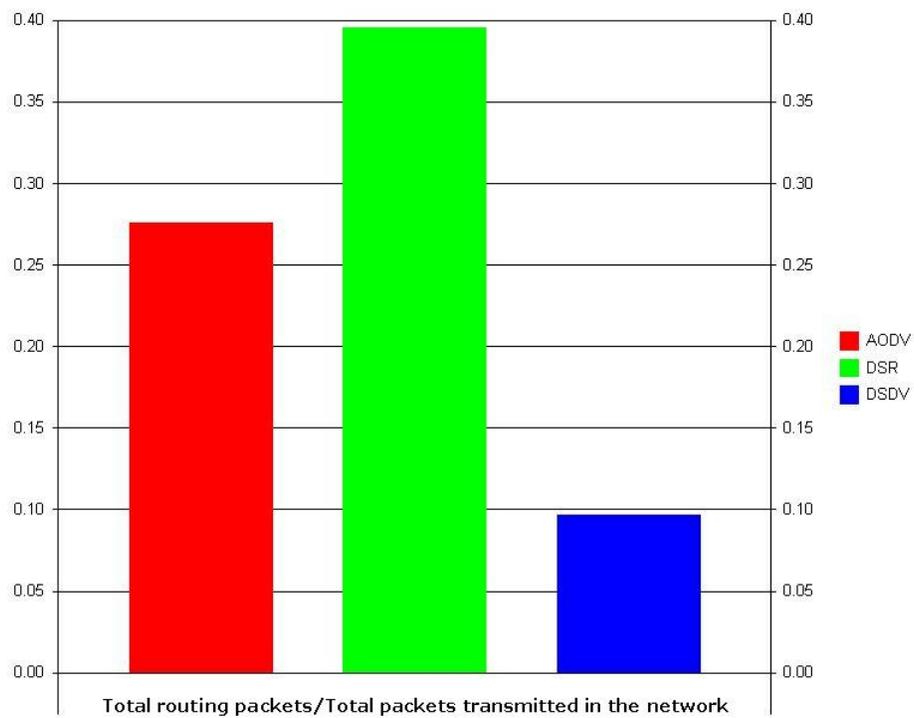

**(c)**

**Figure 9. Routing Load (packets) (a) Scenario 1 (b) Scenario 2 (c) Scenario 3**

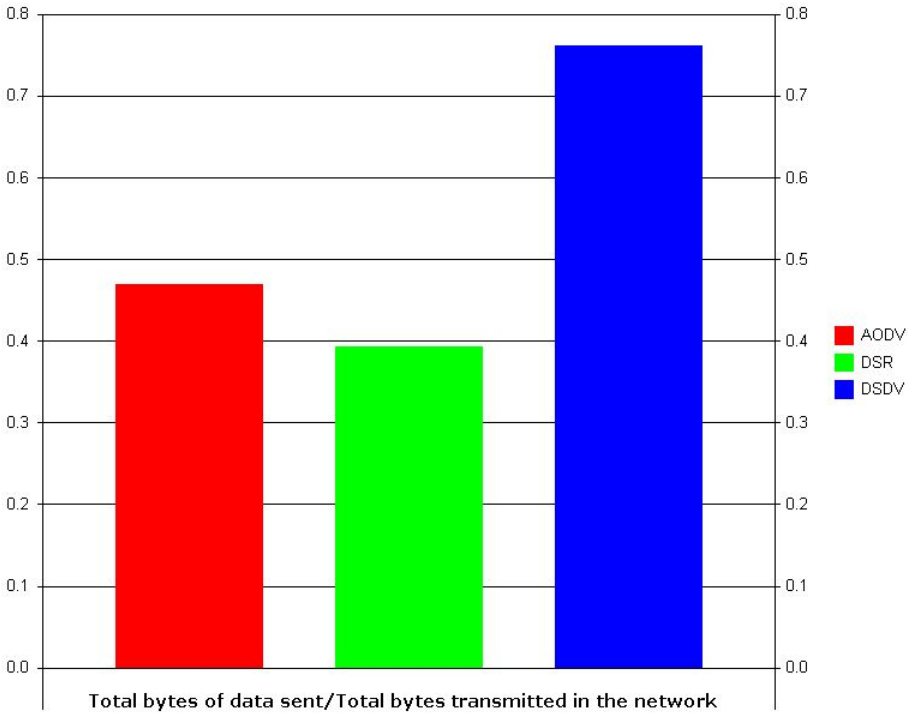

(a)

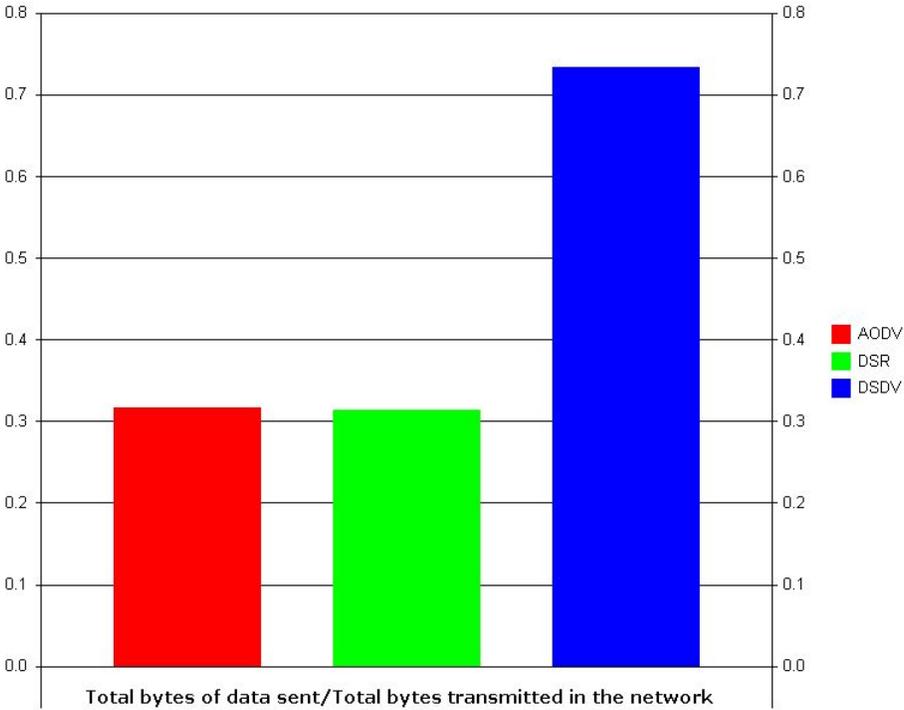

(b)

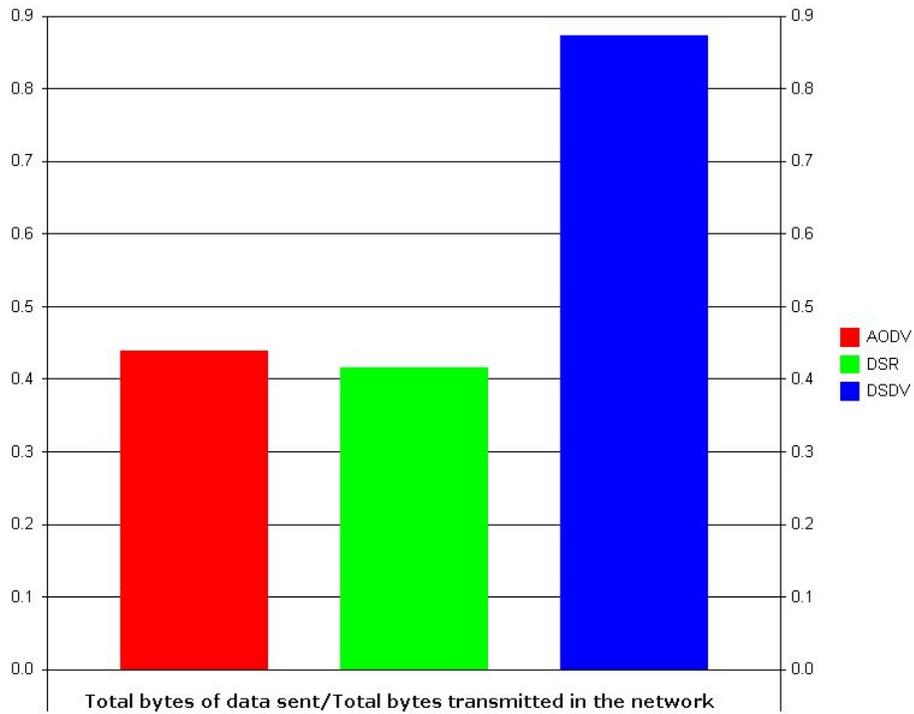

**(c)**
**Figure 10. Goodput (bytes) (a) Scenario 1 (b) Scenario 2 (c) Scenario 3**

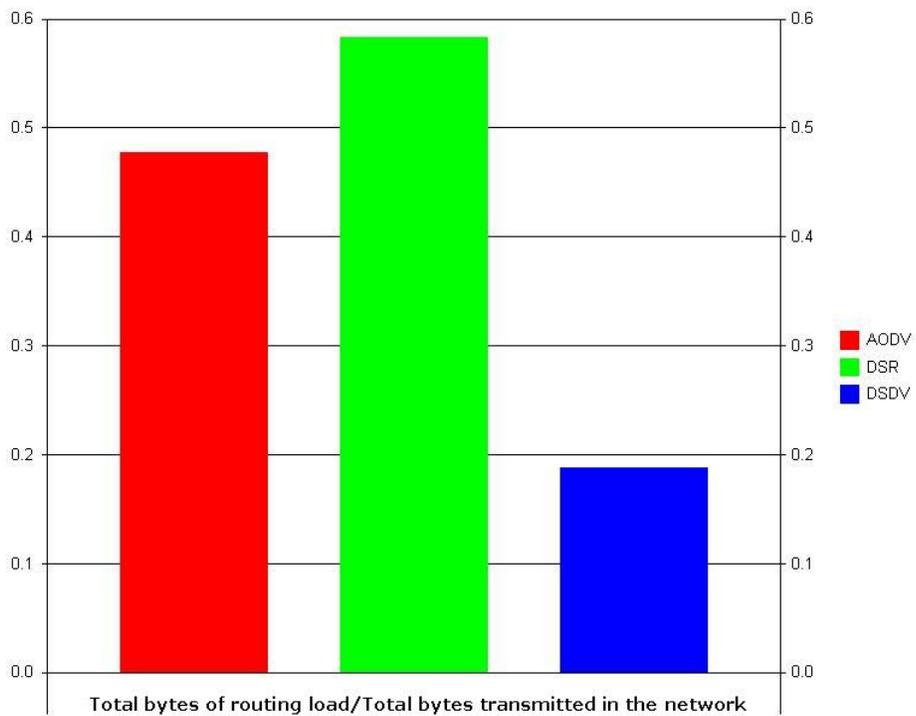

**(a)**

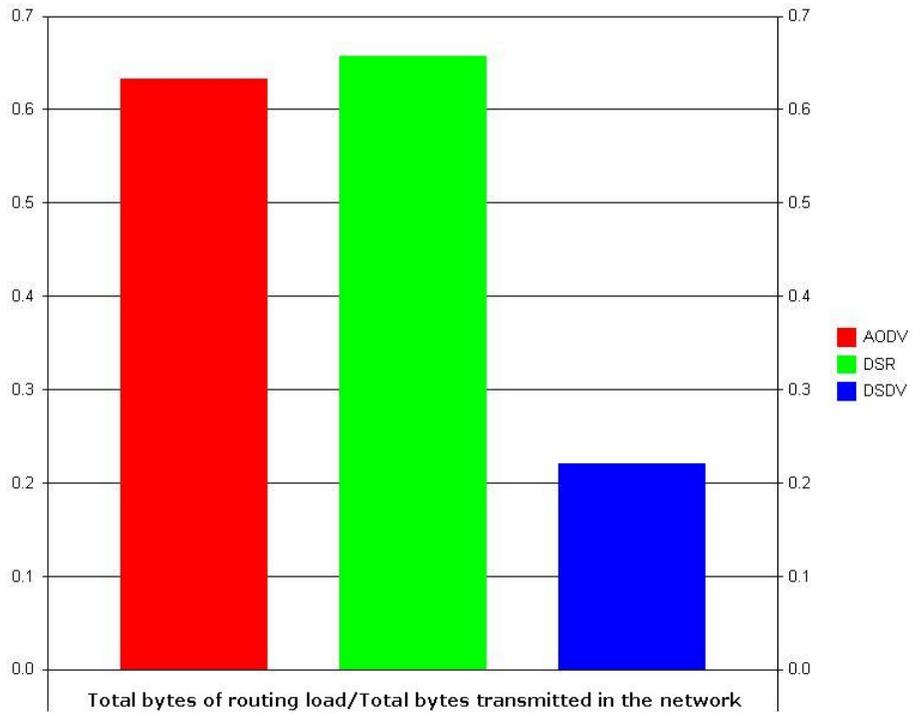

(b)

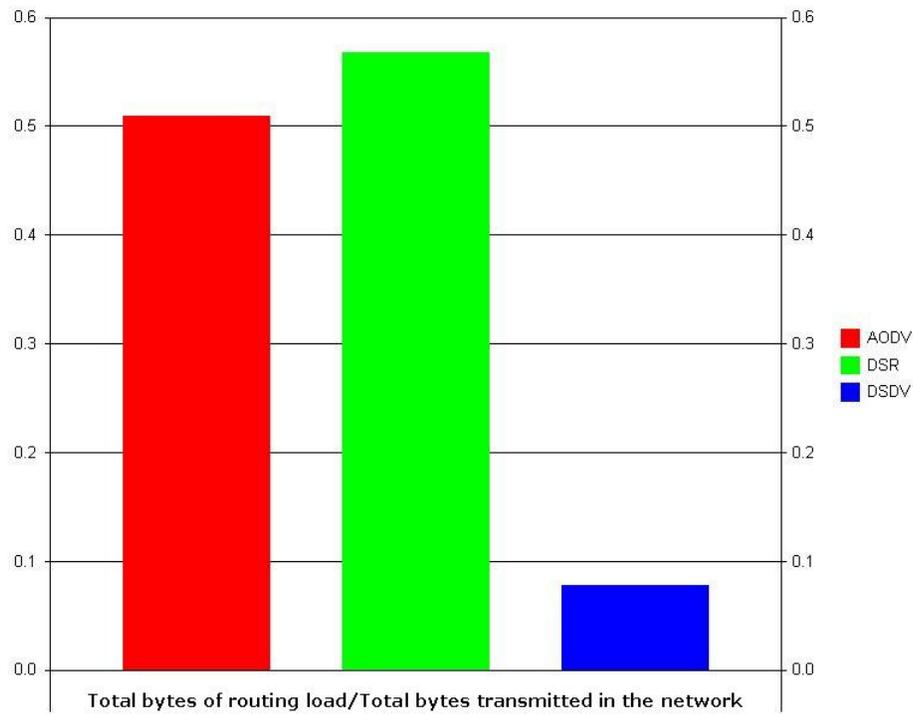

(c)

**Figure 11. Routing Load (bytes) (a) Scenario 1 (b) Scenario 2 (c) Scenario 3**

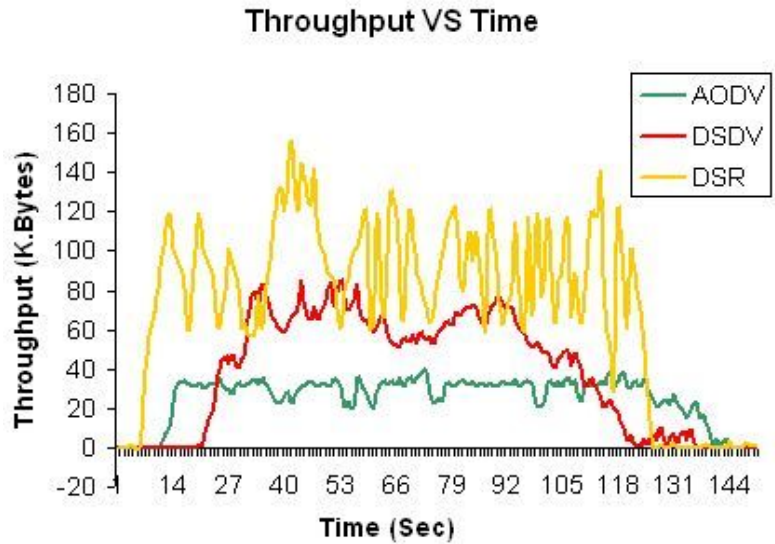

**Figure 12. Throughput vs Time.**

Figure 12 shows the aggregated result for 'Throughput vs Time' where we analyzed the total bytes received by the destination node per second (Data packets and overhead). Based on the results that we see here, the following comments could be made:

**AODV:** starts off quickly and the data rate is more stable.

**DSR:** starts off quickly however we can see that there are lots of fluctuations in the data rate.

**DSDV:** takes time to start off but the data rate has lesser fluctuations.

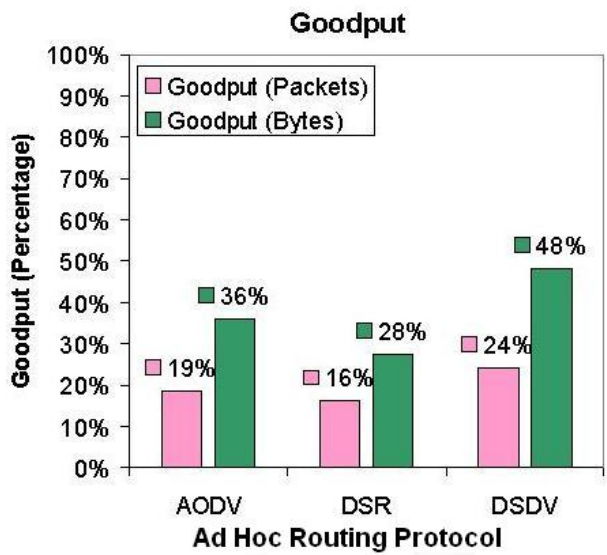

**Figure 13. Goodput for three MANET routing protocols**

We calculated Goodput in terms of number of packets and the packet size in bytes. Now, if we analyze the graph presented in Figure 13, we can see that on an average, if 100 packets are transmitted in the network, 19 packets would be data packets for AODV, 16 for DSR, and 24 for DSDV. In term of bytes, on an average; if 100 bytes of packets are transmitted through the network, 36 bytes would be data packets for AODV, 28 bytes for DSR, and 48 bytes for DSDV. From these data, we could deduce that; though DSDV takes time to converge, it actually is sending more data packets in number as well as in bytes than that of AODV and DSR. Now, the rest of the percentage of each individual graph will be the overheads that contain routing packets and acknowledgements.

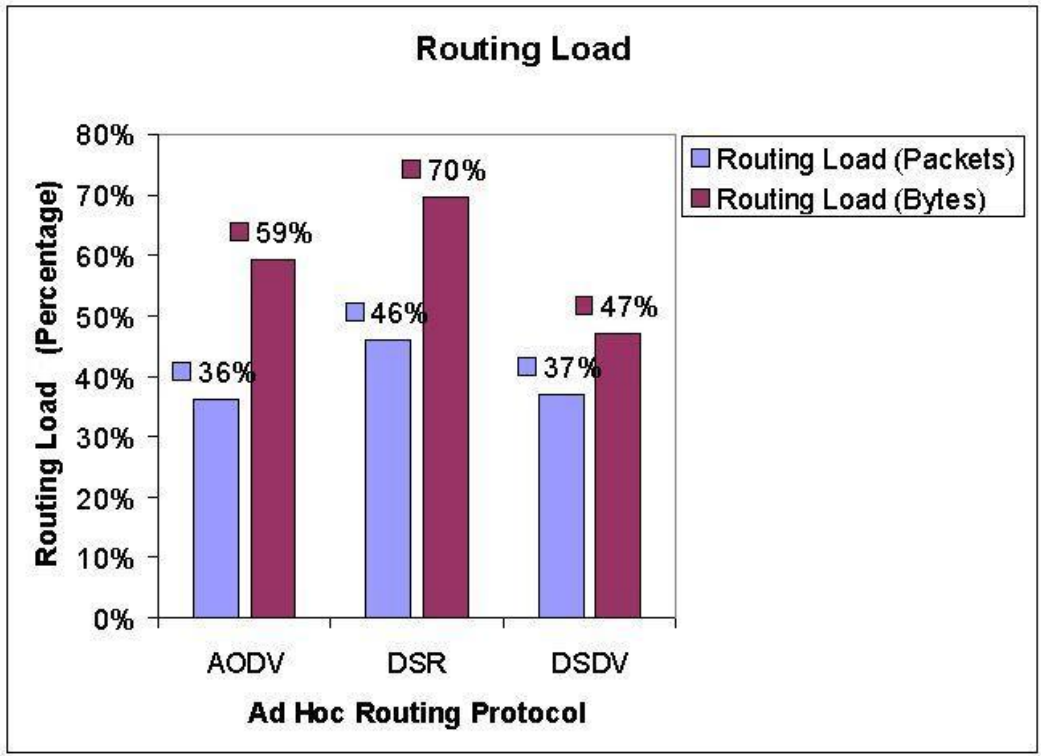

**Figure 14. Routing loads for different experimented MANET protocols**

We again calculated routing loads in terms of number of packets and the packets size in bytes. The results are presented in Figure 14. Again we can see that; though DSR has a better throughput, it actually contains more overhead for routing packets. However, DSDV has a relatively lower routing load than

AODV and DSR.

## 5 RELATED WORKS AND SALIENT FEATURES OF OUR WORK

Luo and Tsai [9] present a graphical simulation system for modeling and analysis of sensor networks. A formal model of sensor networks is proposed based on Space Time Petri Net (STPN). STPN is a formal model language which is extended from Time Petri Nets (TPN) and Colored Petri Nets (CPN). The STPN not only can model the various characteristics of sensor networks, such as temporal and spatial information, but also can simulate the different behaviors of sensor networks. For validating and evaluating the modeling power of the STPN, the authors implement a graphical modeling and simulation environment based on STPN. Using this tool, a designer or architect can model or simulate certain protocols or algorithms for sensor networks and analyze their characteristics and performance. Finally, experimental results of simulation analysis are presented. This work is restricted to a certain type of analysis, needs some advanced knowledge about petri nets, and the prototype that is developed is not user-friendly.

Rosen et al. [10] describe a simulation tool, SHOPMET; used to study the creation and optimization of propagation maps for Node State Routing protocols within wireless ad hoc networks. The simulation tool is developed using MATLAB and interfaces with DLL libraries for network data management support. These DLL libraries also have applications within OPNET [11] for other Node State Routing studies. Propagation maps are used in MANET routing protocol Node State Routing (NSR) to capture how well a node can hear its neighbors. SHOPMET has been designed to investigate the best algorithm to deploy NSR as well as the methodology for collecting propagation observations that best suits NSR. It provides plotting of propagation maps and network animation to demonstrate the propagation map optimization algorithms in various network environments. In addition, it also contains technical solvers to perform experimentation on the algorithm. Though the authors have tried to develop an all-in-one tool, this is too focused to work with a specific type of routing protocol and thus the scope of using it for various purposes is limited.

In [12], authors present an emulation platform called NEMAN that allows running a virtual wireless network of hundreds of nodes on a single end-user machine. NEMAN is shown to be an important and useful tool during development of different applications and protocols for the authors' project, including a key management protocol and a distributed event notification service. It should be noted that NEMAN is an emulator, not a simulation tool. However, we mention this as it includes a Graphical User Interface (GUI) to visualize the emulated network.

In [13], the authors present their interactive ns-2 protocol and environment confirmation tool (iNSpect). The iNSpect program was developed to allow direct visualization and analysis of ns-2 based wireless simulations. iNSpect is a lively tool that can be utilized with minimal overhead and The iNSpect program uses a GTK+ graphical user interface (GUI) for direct scene manipulation. The iNSpect program is multi-platform and executes in Linux, MacOS X, Windows, and Cygwin. In addition to ns-2, this tool can also be used with any simulator or testbed environment which produces output in the iNSpect expected file format. Though iNSpect on the surface seems to be an elegant tool for visual simulation, it is not user-friendly and requires much effort to make the codes run properly.

Lei Shu et al. [14] present NetTopo; an integrated java-based framework that associates with both simulation and visualization functions of real testbed scenarios in order to assist the investigation of different algorithms in Wireless Sensor Networks (WSNs). NetTopo exhibits the simulation of virtual WSN, the visualization of real testbed, and the interaction between simulated WSN and testbed effectively, corresponding to large-scale wireless sensor networks to provide an accurate result where it has been deployed in a realistic environment such as monitoring an erupting Volcano, rescuing in a sudden earthquake or monitoring hazardous situations. This tool also fails to provide much flexibility to the user and cannot simulate the required routing scenarios in WSN. Other than these works, some good surveys on different simulation tools for MANET (or any other wireless networks) could be found in [15] and [16].

There are many works like [17, 18, 19, 20] available which have evaluated the performances of different MANET routing protocols. However, in our work we used our own developed tool to simplify the tasks of multiple simulations at a time. We also have considered various scenarios to run our simulations. We already have talked about various simulation tools and their features. Compared to all other works, ViSim gives much more user-friendliness, flexibility, functionalities, and ease of using complex simulation scenarios. The main difference between ViSim and any other simulation tool is that ViSim uses ns-2 simulations in the background but makes the tasks lot easier for the users. The plotting of graphs is also a great feature and this facility was not included in most of the previously developed graphical simulation tools. Our tool is developed in such a way that it can be used by all types of users. A naive user can use it for visualizing simulations or for simulation demonstration. An expert user can write his own tcl script and run it using ViSim tool and then can generate various performance comparison graphs without taking resort to other cumbersome programming methods.

## 6 CONCLUSIONS AND FUTURE WORKS

In this paper, we have presented our user-friendly simulation tool/prototype which can ease the task of simulation of MANET routing protocols even in Windows based environments. Many users dealing with ns-2 simulations face troubles in setting up Linux or other systems and environment. The use of ActiveTcl with graphical ViSim interface could really be beneficial for the research community in general. Using our simulation tool, we obtained different graphs and analyzed the results for different scenarios which lead us to the following conclusions:

1. For AODV, we can see that it adapts quickly to the change of the network and has a relatively stable throughput with a moderate goodput. So, in an application where there is a fast change in the network topology and a requirement of stable date rate, AODV is more preferable.

2. DSDV turns out to have the best goodput and lesser routing load; however, it takes time to converge. So if there is relatively less number of nodes in the network and the mobility is somewhat steady or slow, DSDV will work more efficiently.

3. DSR, though has a very high throughput, it actually contains less data packets and we can see that there are lots of fluctuations on the throughput curve which are not preferred in a wireless network.

As our future works, we would like to add more functionalities to ViSim with easy access to the programming codes and parameter changes for various network scenarios. We'd also like to investigate other established routing protocols to make a full-scale comparison using our visual simulation tool, ViSim. For the information about the official release of ViSim version 1.0, the readers are encouraged to visit: http://faculty.bracu.ac.bd/~spathan/research/visim.html

## ACKNOWLEDGMENTS

The authors like to give special thanks to Taufiq Abdur Rahman and Sadia Hamid Kazi for their kind cooperation in doing this work.

## Appendix

Details about the features and functionalities of our simulation tool; ViSim, its installation, technical report, and user's manual can be found in this URL:

http://faculty.bracu.ac.bd/~spathan/research/visim.html .